\documentclass[oneside,article]{memoir}

\usepackage{amsmath}
\usepackage{amssymb}
\usepackage{caption}
\usepackage{geometry}
\usepackage{bm}
\usepackage{multirow}
\usepackage{array}
\usepackage{footmisc}
\usepackage{longtable}
\usepackage{ragged2e,array,longtable}
\usepackage{booktabs} 
\usepackage{hyperref}
\usepackage{physics}
\usepackage[version=4]{mhchem}
\usepackage{siunitx}
\usepackage{float}
\usepackage{textcomp}
\usepackage{mathptmx}
\usepackage{graphicx,import}
\usepackage{tabularx}
\usepackage{xcolor}
\usepackage{dblfloatfix}
\usepackage{fix-cm}
\usepackage{orcidlink}
\usepackage{authblk}
\usepackage[normalem]{ulem}
\usepackage[backend=biber,style=chem-acs,articletitle=true,doi=true]{biblatex}
\setsecnumdepth{subsection}

\newcommand{\manuscripttitle}{From Local Structure to Thermodynamics and Transport of Water with Machine Learning Force Fields
}

\title{\manuscripttitle}
\date{\vspace{-5ex}}

\begin{document}
\emergencystretch=3em

\begin{frontmatter}

{\setlength{\parindent}{0pt}

\author{
Andreas Kretschmer\textsuperscript{*,$\dagger$}\orcidlink{0000-0003-1297-8690},
Florian Altmann\textsuperscript{$\dagger$}\orcidlink{0009-0000-7808-9744},
Nader Nour\orcidlink{0009-0008-8295-3492},
Alper T. Celebi\textsuperscript{*}\orcidlink{0000-0001-7727-194X},
and Markus Valtiner\orcidlink{0000-0001-5410-1067}
}

\date{
\small Institute of Applied Physics, TU Wien, Wiedner Hauptstra{\ss}e 8--10, 1040 Vienna, Austria\\
\small *Corresponding authors: \href{mailto:andreas.kretschmer@tuwien.ac.at}{andreas.kretschmer@tuwien.ac.at}, \href{mailto:celebi@iap.tuwien.ac.at}{celebi@iap.tuwien.ac.at} \\
$\dagger$ These authors contributed equally
}

\maketitle
}

\DoubleSpacing

\begin{abstract}
\noindent
We evaluate machine learning force fields derived from different density functional theory exchange correlation functionals using the full six-dimensional pair correlation function of liquid water, three-body structural descriptors, excess entropy, and transport properties. 
The predicted microscopic structure and dynamics depend strongly on the underlying functional: neglecting dispersion produces pronounced over-structuring, overly negative excess entropy, and suppressed diffusion. 
Translational and orientational entropy contributions are tightly coupled and together exhibit a clear relationship with the reduced self-diffusion coefficient. 
Among the tested models, RPBE-D3 provides the most consistent agreement with experiment across structural, thermodynamic, and transport properties. 
The classical SPC/E model serves as an additional reference and displays notable similarities to RPBE-D3, consistent with the comparable Born effective and partial charges of the two models.
\end{abstract}


\begin{figure}
    \centering
    \includegraphics[width=1\linewidth]{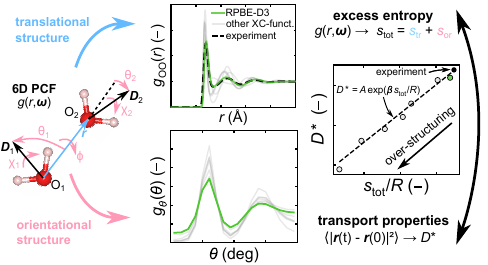}
    \caption*{Graphical abstract}
    \label{fig:graphical_abstract}
\end{figure}

\end{frontmatter}

\begin{mainmatter}
\DoubleSpacing

\section*{Introduction}
Water's molecular structure presents a profound challenge to chemical physics.\autocite{brini_how_2017} The topology and fluctuations of its hydrogen bond network are fundamentally linked to water's hallmark anomalies,\autocite{errington_relationship_2001} including its density maximum, unusual heat capacity, and anomalous diffusion behavior. 
In aqueous electrolytes, this local structure is perturbed in an ion-specific manner,\autocite{shi_impact_2023} thereby shaping solvation environments,\autocite{marcus_effect_2009} adsorption patterns at interfaces,\autocite{olgiati_entropic-dielectric_2026} and ultimately the kinetics of electron- and proton-transfer processes central to electrocatalysis.\autocite{kornyshev_kinetics_2003,carpenter_picosecond_2018} These effects have been directly associated with changes in molecular correlations,\autocite{bandyopadhyay_correlation_2013,nilsson_structural_2015} underscoring the need for models that accurately capture water's local structure. A complete description of water must bridge these microscopic correlations with macroscopic thermodynamic and dynamical properties. In this sense, excess entropy provides such a thermodynamic link between the translational packing of molecules and orientational organization of the hydrogen bond network with macroscopic liquid properties. Transport properties such as the self-diffusion coefficient and viscosity provide complementary dynamical benchmarks, since they reflect the collective rearrangement of the hydrogen bond network\autocite{ohmine_water_1999,gomez_water_2022} and control mass transport, mixing,\autocite{rolle_mixing_2019} and reaction rates\autocite{collins_diffusion-controlled_1949} in aqueous environments. Consequently, an accurate description of molecular correlations, excess entropy, and transport is essential for linking atomistic structure to liquid water properties and electrochemical function.

\begin{figure}
    \centering
    \includegraphics[width=0.8\linewidth]{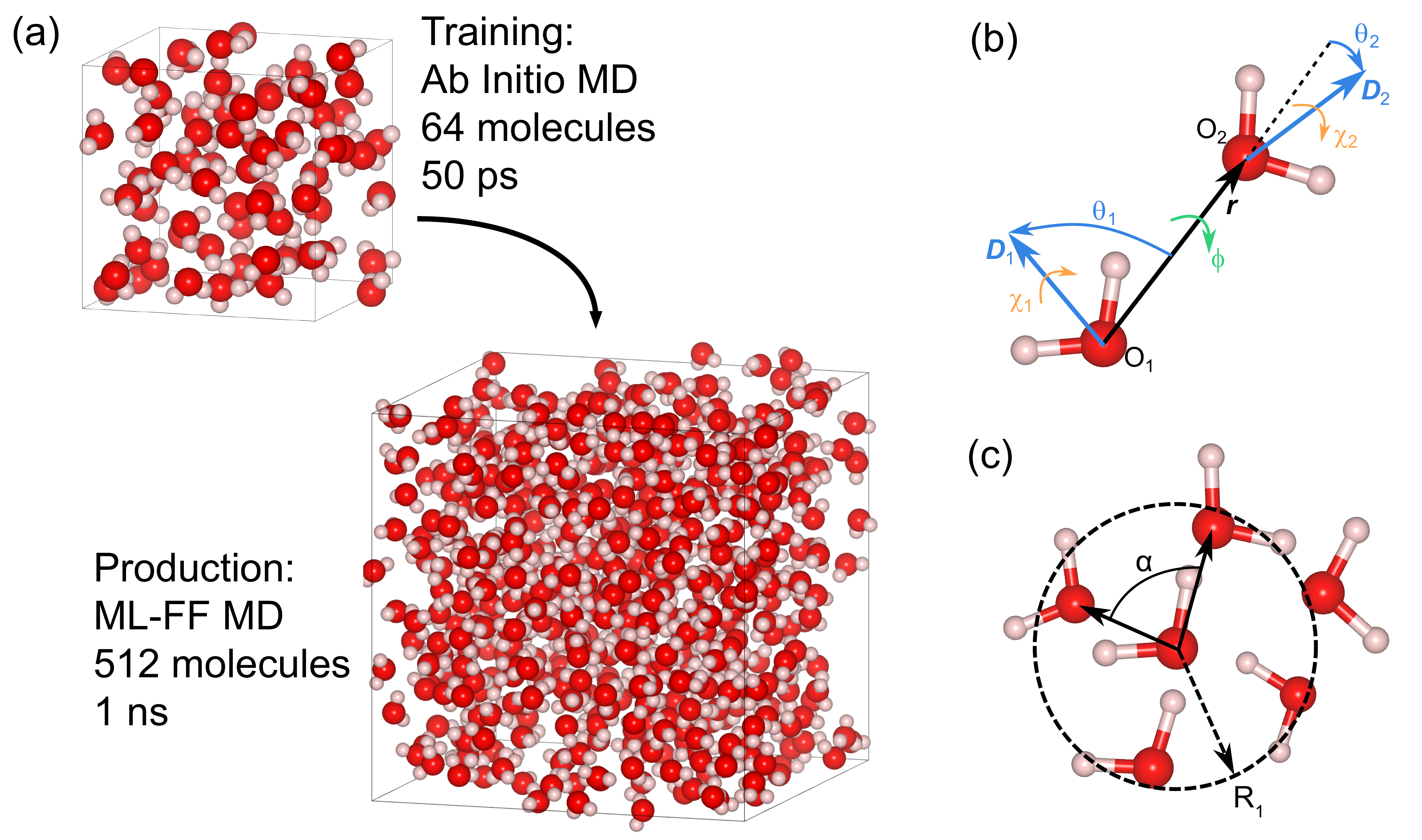}
    \caption{(a) Sketch of the simulation workflow with seven different XC-functionals. (b) Definition of the molecular PCF of water, the angles and vectors are defined in the methods section. \autocite{lazaridisOrientational1996} 
    (c) Definition of the three-body-angle $\alpha$ between two vectors connecting a central water oxygen atom to oxygen atoms of two neighboring water molecules. For the computation of the three-body-angle distribution, all angles formed by unique pairs of water molecules within the first hydration shell of radius $R_1$ are considered.}
    \label{fig:Workflow_Definition}
\end{figure}

Molecular simulations provide a direct route to resolve these correlations at atomistic resolution, however, their predictive power depends critically on the underlying interaction potential. 
Classical force fields, while computationally efficient and applicable over large length and time scales, are constrained by fixed functional forms and empirical parameterizations \autocite{demerdash_advanced_2018,dick_chapter_2005,kadaoluwa_pathirannahalage_systematic_2021}. 
As a result, they often fail to fully capture subtle many-body effects, polarization, and interfacial restructuring. 
Machine learning force fields (ML-FFs) have emerged as a powerful alternative, offering near Density Functional Theory (DFT) accuracy with a computational cost that remains accessible to large-scale molecular simulations that are required for this purpose \autocite{schran_machine_2021,monterodehijesComparing2024,omranpour2024,liu_toward_2022}. 
However, it remains unclear how the choice of exchange correlation (XC) functionals used to generate the DFT training data propagates into the molecular structure and various physical properties predicted by ML-FFs. 
To this goal, we systematically evaluate ML-FFs obtained from seven different XC descriptions of liquid water: PBE \autocite{perdew1996}, PBE-D3 with zero damping \autocite{perdew1996,grimme2010}, PBE-TS \autocite{perdew1996,tkatchenko2009}, RPBE \autocite{hammer1999}, RPBE-D3 with zero damping \autocite{hammer1999,grimme2010}, R2SCAN+rVV10 \autocite{ning2022}, and vdW-DF-cx \autocite{berland2014}. 
For each XC-functional, a separate ML-FF was trained using a simulation cell containing 64 water molecules and the on-the-fly learning scheme implemented in the Vienna Ab initio Simulation Package (VASP). \autocite{jinnouchi2020} 
The resulting ML-FFs were subsequently employed in production simulations. 
Starting from equilibrated NpT simulations at \SI{1}{bar} and \SI{300}{K} to determine the liquid density, we analyzed NVT trajectories containing 512 water molecules using the full six-dimensional molecular pair correlation function (PCF) of water\autocite{lazaridisOrientational1996}, three-body angular correlations, and hydrogen bonding characteristics. 
The overall workflow is illustrated in \autoref{fig:Workflow_Definition}(a), while the molecular PCF and the three-body angle are defined in \autoref{fig:Workflow_Definition}(b) and (c), respectively.
These structural descriptors were then related to the excess entropy, self-diffusion coefficient, and viscosity, allowing us to trace how the underlying XC-functional propagates from molecular correlations to thermodynamic and transport properties. 
Throughout this work, the resulting ML-FF predictions are benchmarked against the well-established classical SPC/E water model.\autocite{berendsen_missing_1987}

\section*{Results \& Discussion}
\subsection*{Model Generation \& Density Validation}
\begin{figure}
    \centering
    \includegraphics[width=\linewidth]{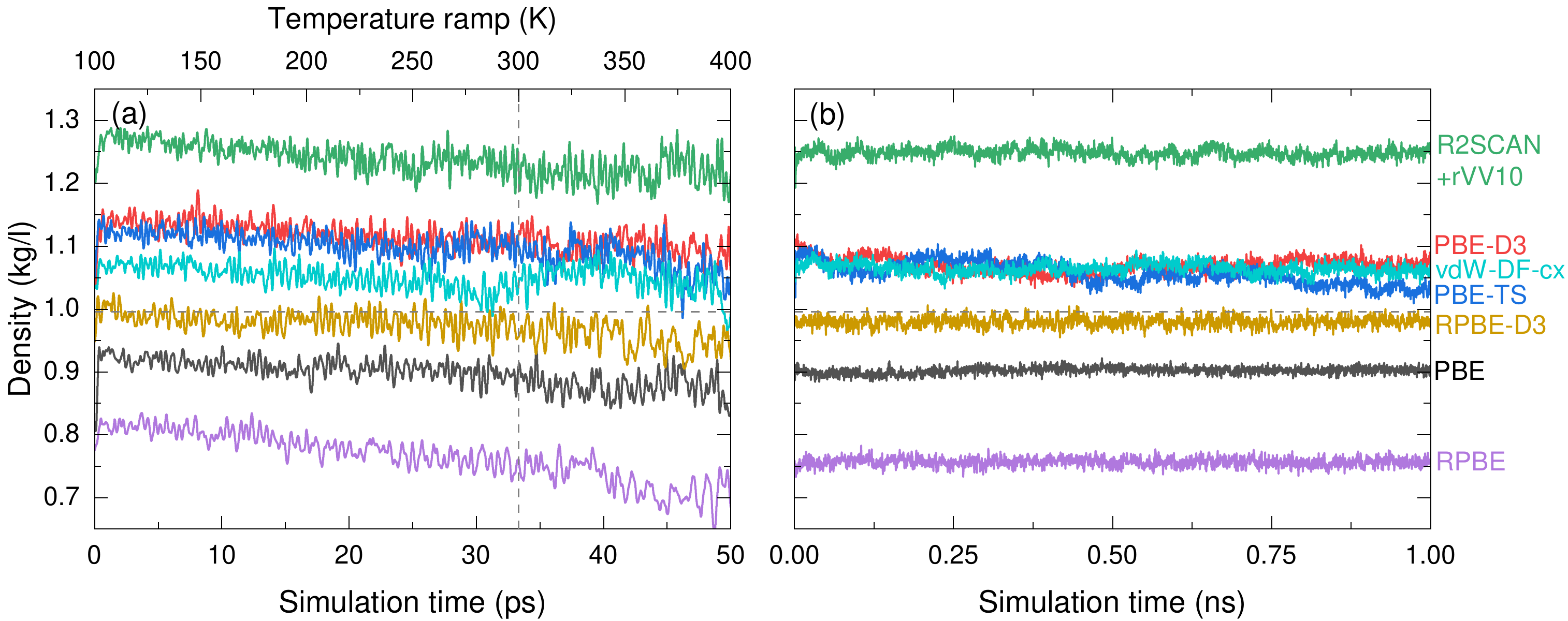}
    \caption{(a) Predicted densities along the training trajectory. The dashed lines indicate the experimental density at room temperature. (b) Variation of densities at \SI{300}{K} and \SI{1}{\bar} during \SI{1}{ns} NpT-ML-FF simulations. The dashed line shows the experimental density.}
    \label{fig:Density}
\end{figure}

As a first validation of the generated ML-FFs, we examined their predictions of the liquid density.
\autoref{fig:Density}(a) shows the evolution of the water density during the NpT training simulations performed for the different XC-functionals.
Here, the initial density was set to \SI{1}{\kilogram\per\liter} for all XC-functionals and then pre-equilibrated for \SI{25}{ps} in the NpT ensemble at \SI{1}{\bar} to avoid large density fluctuations during training, which would increase the errors in the ML-FFs.
The corresponding training errors, basis set size, temperature $T$, energy $E$, volume $V$, and pressure $p$ are shown in the Supporting Information for each XC-functional. The training was performed along a heating ramp from \num{100} to \SI{400}{K} to access a wider phase space of local configurations, leading to a decreasing density with increasing temperature as expected. No phase transition or anomalous density maximum are evident from this heating ramp, which would require long equilibration times instead \autocite{singraberDensity2018}.

It is well established that the choice of XC-functional significantly impacts the predicted density.\autocite{gillan_perspective_2016} Specifically, RPBE and R2SCAN+rVV10 result in the greatest underestimation and overestimation in comparison with the experimental value (see dashed line in \autoref{fig:Density}(a)), respectively. 
The inclusion of D3 and TS dispersion corrections increases the predicted density and brings RPBE-D3 into close agreement with the experimental value, whereas the corrections lead to overcompensation in the case of PBE. 
Notably, RPBE-D3 provides the most accurate density prediction, with a deviation of less than \SI{3}{\percent}. 
The PBE-D3, PBE-TS, and vdW-DF-cx XC-functionals all overestimate the density, with vdW-DF-cx exhibiting the smallest error, followed by PBE-TS and PBE-D3.

Importantly, the densities predicted by the trained ML-FFs remain stable over extended timescales in the production simulations. 
\autoref{fig:Density}(b) shows the evolution of the density during \SI{1}{ns} NpT simulations performed at \SI{1}{\bar} and \SI{300}{K}.
The resulting equilibrium densities were subsequently used to define the simulation cells for the NVT simulations employed in the analysis of structural and dynamical properties. 
Notably, the density can be corrected to the experimental value either by applying an external pressure (NVT or NpT ensemble) and/or significantly higher cut-off energies in the NpT ensemble \autocite{monterodehijesComparing2024}.
However, we are interested in isobaric conditions and the resulting impact on the properties. 
The corrections would preclude the use of the ML-FF in later applications like solid/liquid interfaces by affecting the interfacial structure due to the pressure, or intolerable computational cost. 
We instead choose to explore these force fields at their native predicted densities. 
Interestingly, all XC-functionals yield qualitatively similar electronic densities of states at the end of the training trajectories as shown in the Supporting Information. 
While the band gaps (see Supporting Information) are strongly underestimated compared to experiment in all cases, this deficiency does not appear to translate directly into the predicted liquid densities, which are instead governed by the balance of intermolecular interactions encoded in the trained force fields.

\subsection*{Molecular Structure from Pair Correlations}
\begin{figure}
    \centering
    \includegraphics[width=\linewidth]{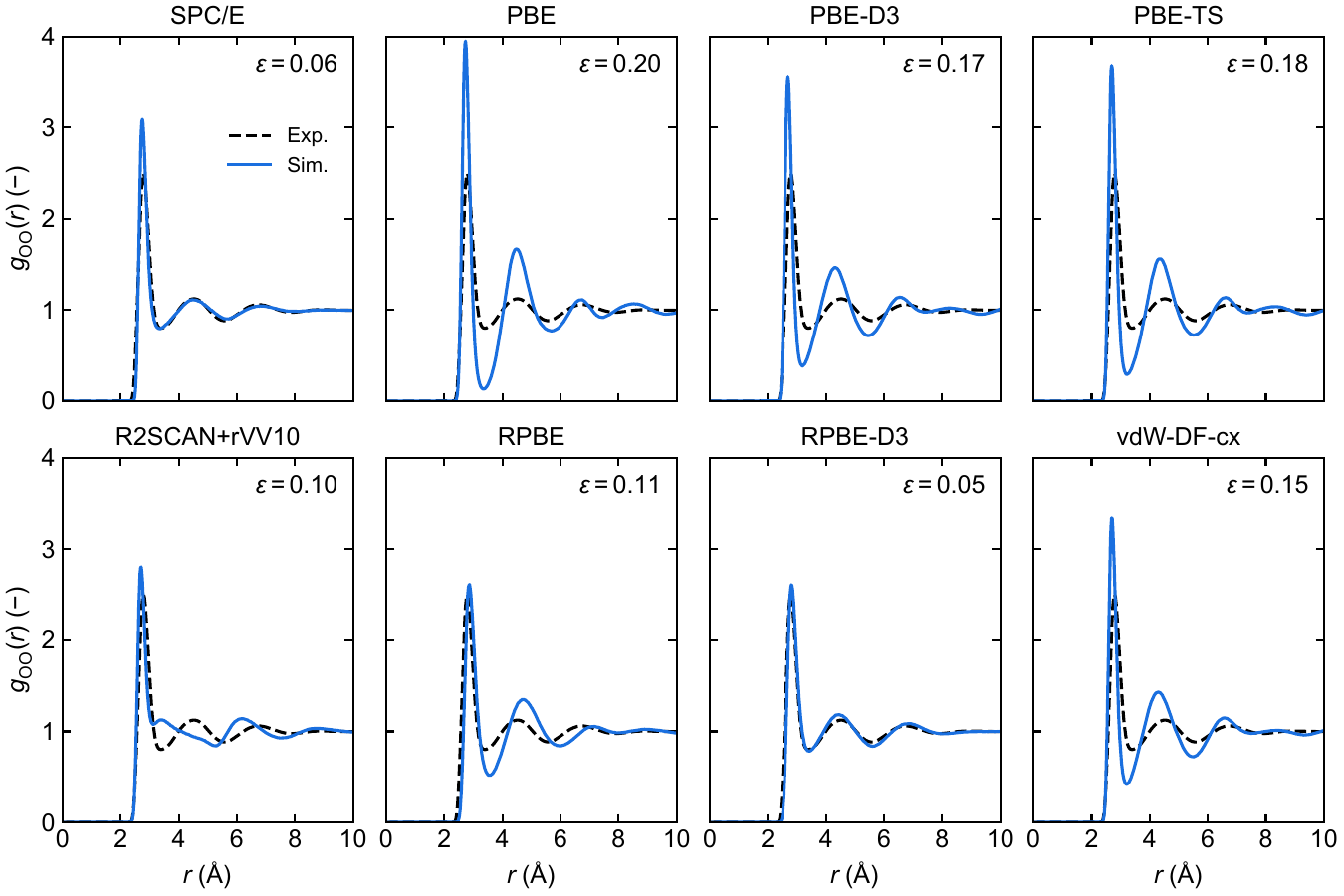}
    \caption{Comparison of oxygen--oxygen RDFs $g_\mathrm{OO}(r)$ obtained from the classical SPC/E model and the ML-FFs based on different XC-functionals with experimental data from Soper \emph{et al.}~\autocite{soperRadial2013}. The value of $\varepsilon$ reported in the top-right corner denotes the root-mean-square error between the MD-derived and experimental RDFs.}
    \label{fig:RDF-OO}
\end{figure}

To assess how the different force fields affect the molecular structure of liquid water beyond simple density variations, we evaluate the full six-dimensional molecular PCF\autocite{lazaridisOrientational1996} $g(r,\vb{\omega})$, defined in the \autoref{fig:Workflow_Definition}(b), from the equilibrated NVT trajectories. The oxygen--oxygen radial distribution functions (RDFs) $g_\mathrm{OO}(r) \equiv g(r)$ obtained for the different XC-functionals and the classical SPC/E model are shown in \autoref{fig:RDF-OO}. These RDFs correspond to the radial projection of the full molecular PCF, obtained by averaging over all orientational degrees of freedom. 
Experimental data from Soper \emph{et al.} \autocite{soperRadial2013} are shown in black dashed line, and the root-mean-square error $\varepsilon$ with respect to the experimental RDF is noted in the top right corner. Overall, SPC/E agrees well with experiment, with a slightly overestimated first maximum, which is notable given that the model was primarily parameterized to reproduce the heat of vaporization \autocite{berendsen_missing_1987}.
While the maxima and minima are placed at the correct distances, PBE leads to significant over-structuring, i.e. a larger number of water molecules around a water molecule in the first and second hydration shell with increased H-bond networks similar to ice, evident by the larger maxima and lower minima in the first and second hydration layers. Adding the D3 or TS dispersion correction reduces this over-structuring significantly, but at the cost of shifting the distribution to smaller distances, confirming the higher density observed in \autoref{fig:Density}. 
The results look very similar to the ones obtained using the vdW-DF-cx functional. 
RPBE reproduces the first hydration shell properly, however fails to capture the position and magnitude of the second shell. This is because the second hydration shell is more sensitive to long-range dispersion interactions than the first shell.
Adding the D3 correction therefore yields excellent agreement of $g_\mathrm{OO}(r)$ over the whole radial distance, capturing both the magnitudes and locations of the first, second and third hydration shell. Ultimately, RPBE-D3 yields the best agreement of all tested XC-functionals with experiment, which is also evident from the smallest root-mean-square error between simulated and experimental RDFs.
R2SCAN+rVV10 completely fails to predict the water structure and does not show the formation of distinct hydration layers beyond the first shell. The oxygen--hydrogen and hydrogen--hydrogen RDFs, shown in the Supporting Information, exhibit similar trends where RPBE-D3 shows the best agreement with experiment, indicating the most accurate description of the translational structure of liquid water.

\begin{figure}
    \centering
    \includegraphics[width=1\linewidth]{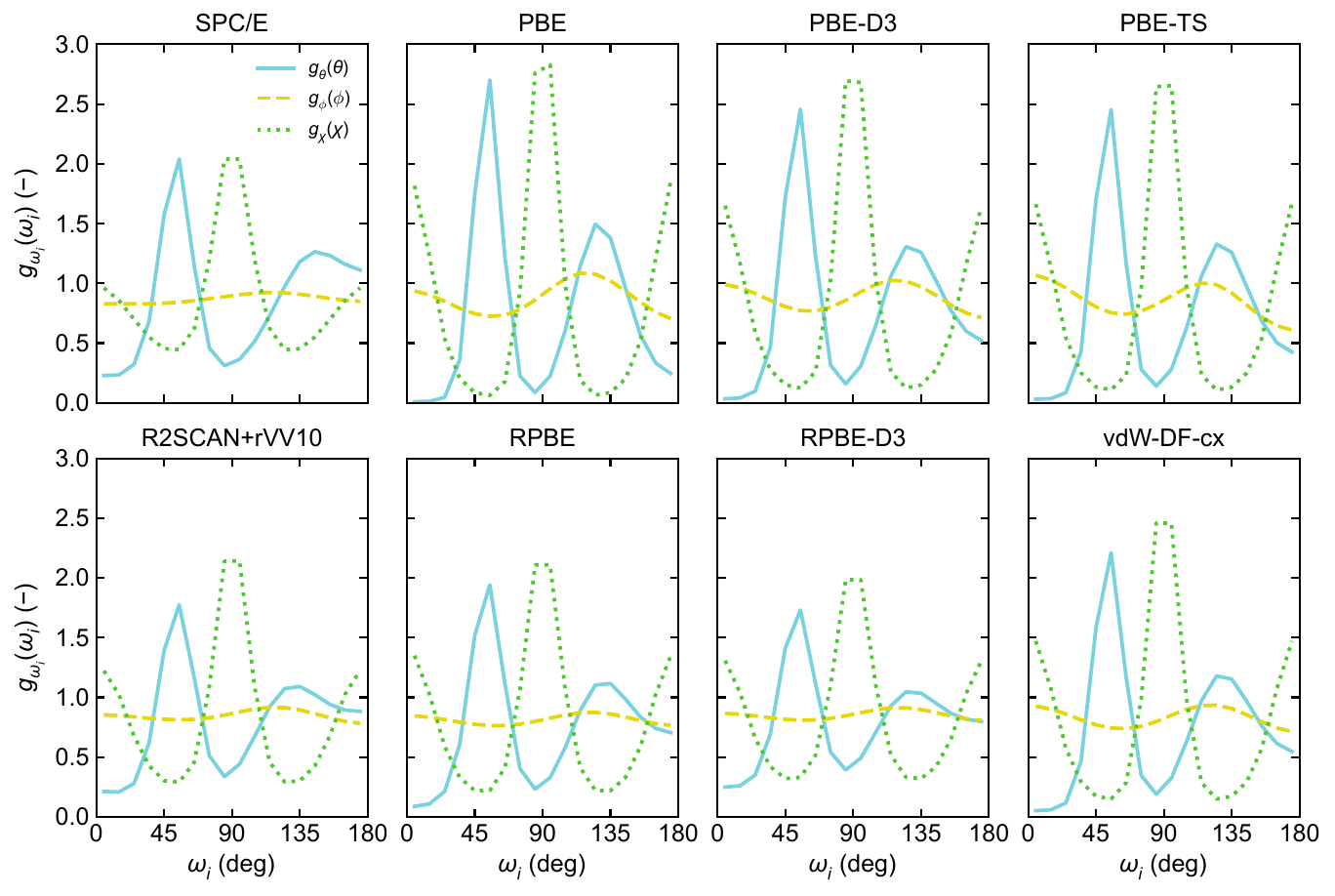}
    \caption{Orientational water structure predicted by the classical SPC/E model and the ML-FFs based on different XC-functionals, represented by the marginal orientational distribution functions $g_{\omega_i}(\omega_i)$.}
    \label{fig:ODF}
\end{figure}

Since water is a polar fluid, orientational correlations dominate over translational correlations.\autocite{lazaridis_solvent_2000} The predicted orientational structure of the different force fields, expressed via the marginal orientational distribution functions (ODFs) $g_{\omega_i}(\omega_i)$, are shown in \autoref{fig:ODF}. Here, the marginal ODFs were computed for molecular pairs within the first hydration shell, defined by the first minimum of \(g_{\mathrm{OO}}(r)\). 
Qualitatively, all force fields yield similar orientational distribution functions, differing primarily in the magnitudes of the respective maxima and minima. This indicates that the principal hydrogen bond structure is consistently captured across all models. This is particularly evident in the angular distribution functions \( g_\theta(\theta) \) (\(\theta_1\) or \(\theta_2\)). The sharp maximum at approximately \SI{50}{\degree} corresponds to the tetrahedral hydrogen bond donor configuration, while the broader maximum around \SI{135}{\degree} is associated with hydrogen bond acceptor configurations \autocite{lazaridisOrientational1996}. 
All force fields reproduce these characteristic features; however, they differ in the probability of alternative configurations. In particular, the class of PBE-based functionals permits almost no configurations for \(\theta < \SI{30}{\degree}\) and strongly suppresses configurations between the two maxima, indicative of an almost ideal tetrahedral structure, which further indicates over-structuring. In contrast, SPC/E and RPBE-D3 allow a finite population of intermediate configurations, reflecting a broader deviation from perfect tetrahedral ordering. 
A similar over-structuring is also observed in the \( g_\chi(\chi) \) (\(\chi_1\) or \(\chi_2\)) ODF. The PBE-based functionals allow pronounced features only at the respective maxima, whereas SPC/E and RPBE-D3 again permit additional configurations outside these preferred orientations. Generally, the D3 and TS correction flatten the $g_\chi(\chi)$ ODF, reflecting increased orientational disorder. 
The $g_\phi(\phi)$ marginal ODF is essentially flat for all the force fields, indicating an absence of a strong angular preference. 
Only the PBE-based functionals exhibit slight angular preferences, suggesting an additional but weak contribution to overall structuring of water.

\subsection*{Tetrahedral and Disordered Motifs in the Hydrogen Bond Network}
\begin{figure}
    \centering
    \includegraphics[width=1\linewidth]{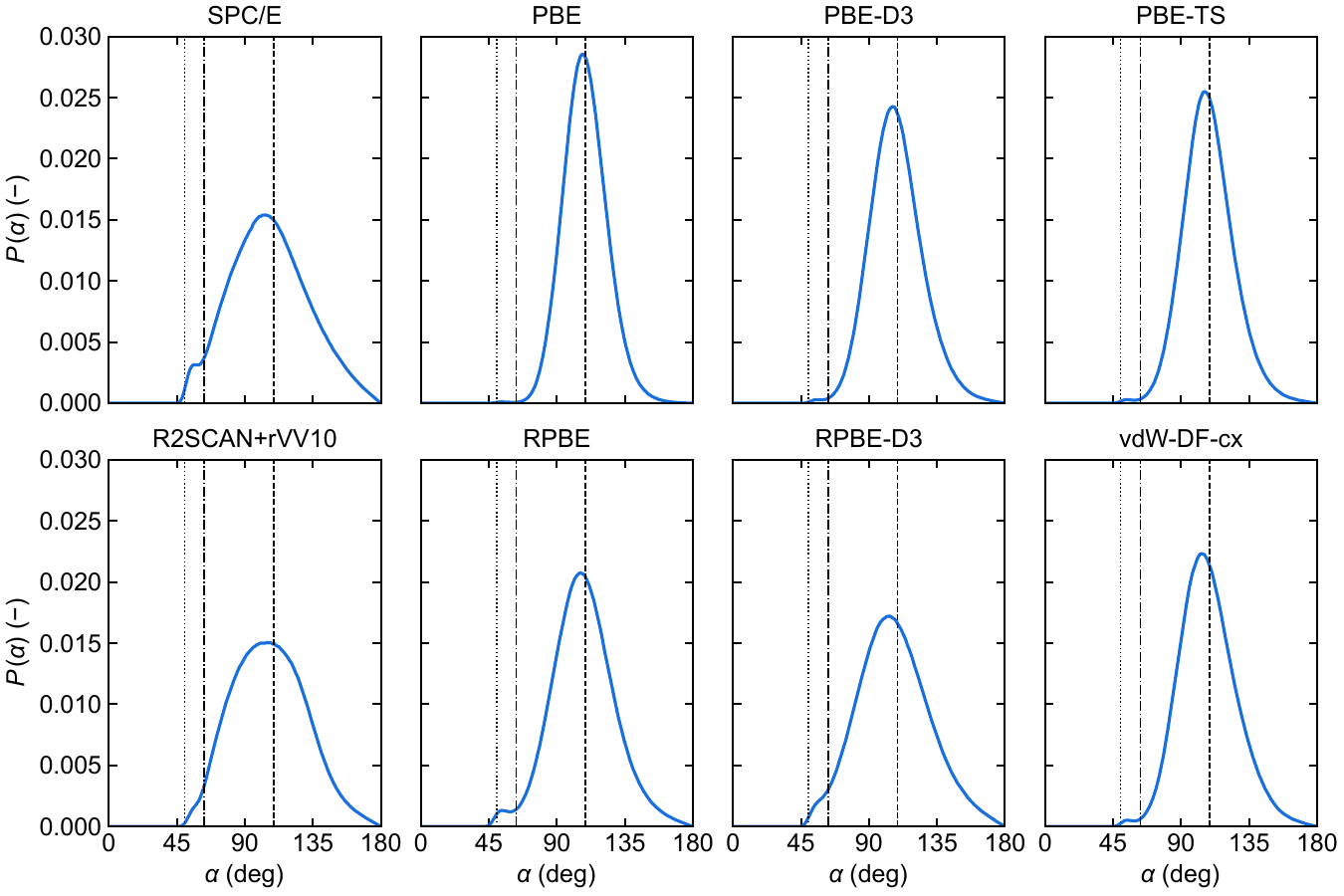}
    \caption{Comparison of the three-body angle distributions obtained from the classical SPC/E and the ML-FFs based on different XC-functionals. The vertical lines indicate characteristic configurations discussed in Monroe \emph{et al.}\autocite{monroe_decoding_2019}: (dashed) perfect tetrahedral alignment at \SI{109.4}{\degree}, (dash--dotted) Lennard-Jones water model at \SI{63.4}{\degree}, and (dotted) interstitial water at \SI{50}{\degree}.}
    \label{fig:ThreeBodyAngle}
\end{figure}

In addition to the molecular PCFs, we analyze the three-body structure, defined in \autoref{fig:Workflow_Definition}(c), of the classical SPC/E and ML-FFs. 
The results are shown in \autoref{fig:ThreeBodyAngle}. 
All distributions are dominated by a broad maximum between \SI{90}{\degree} and \SI{135}{\degree}, which is associated with tetrahedrally coordinated water. 
In a perfect tetrahedral crystalline arrangement, where each water molecule is surrounded by four nearest neighbors in the first hydration shell, one would expect a single, sharp peak at \SI{109.5}{\degree}. 
In SPC/E, an additional smaller peak between \SI{50}{\degree} and \SI{65}{\degree} appears; it corresponds to a fifth neighbor within the first hydration shell occupying an interstitial position in the predominantly tetrahedral network. 
Such a feature becomes dominant for a Lennard--Jones water model without polarity\autocite{monroe_decoding_2019}. 
The relative intensity of these two peaks serves as an indicator of structural order versus disorder in the liquid: the tetrahedral component reflects ice-like ordering due to hydrogen bonding, whereas interstitial water molecules disrupt this order. 
Consistent with the ODF-based hydrogen bond analysis, PBE yields a highly over-structured liquid, evidenced by a tall and narrow tetrahedral peak and no interstitial contribution. 
Here, the D3 and TS corrections introduce disorder only almost imperceptibly, with interstitial populations remaining substantially smaller than in SPC/E.
The vdW-DF-cx and RPBE XC-functionals exhibit excessive tetrahedral structure similar to PBE. 
However, the inclusion of the D3 correction for RPBE yields a much more pronounced population of interstitial and non H-bonded water molecules similar to the SPC/E, effectively reducing the tetrahedral behavior. 
R2SCAN+rVV10 shows a comparatively broad tetrahedral peak, indicating a higher degree of disorder, which arises from an incorrect radial water structure, as observed in the oxygen--oxygen RDF in \autoref{fig:RDF-OO}. 

\begin{figure}
    \centering
    \includegraphics[width=1\linewidth]{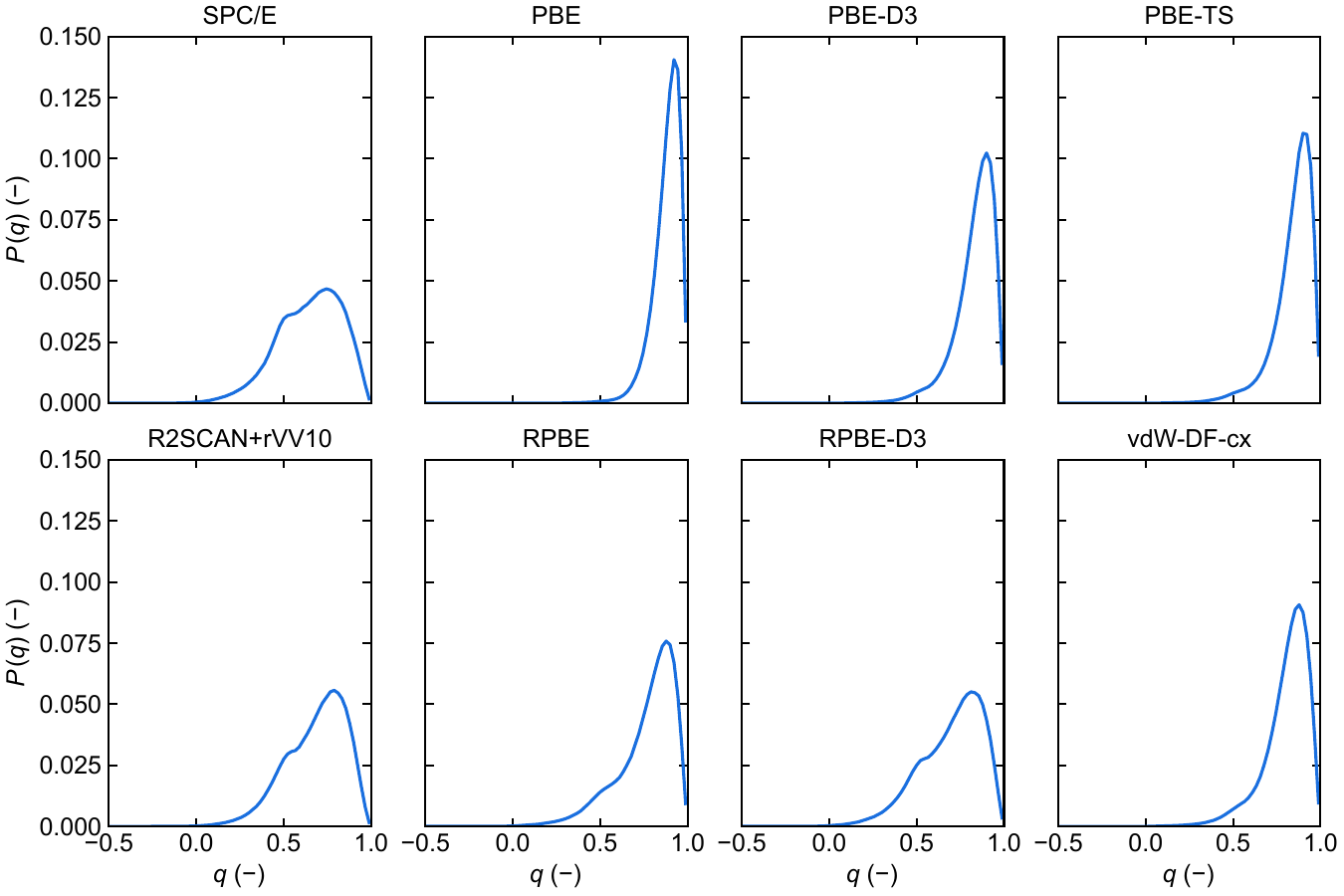}
    \caption{Comparison of the distribution of the tetrahedral order parameter $q$ from the classical SPC/E and the ML-FFs. } 
    \label{fig:qFactor}
\end{figure}

The coexistence of ordered and disordered local water environments within the first hydration shell is further reflected in the probability distribution of the tetrahedral order parameter $q$, shown in \autoref{fig:qFactor}.
Here, the classical SPC/E force field as well as the R2SCAN+rVV10 and RPBE-D3 XC-functionals exhibit bimodal $q$ distributions with similar intensity, indicating that the local arrangement of a central molecule and its four nearest neighbors populates two distinct motifs: an \emph{ice}-like (high-$q$) configuration and a more disordered (low-$q$) configuration. 
This interpretation is consistent with the respective three-body angle distributions in \autoref{fig:ThreeBodyAngle}. 
For the SPC/E, R2SCAN+rVV10, and RPBE-D3 force fields, we observe both, tetrahedral alignment representing \emph{ice}-like configurations, and a substantial population of interstitial arrangements, representing disordered configurations. 
No bimodality is observed for the class of PBE XC-functionals, similar to the distribution produced by vdW-DF-cx.
The RPBE XC-functional also yields predominantly ordered configurations, but to a lesser extent, forming a suitable basis for the D3 correction.

Overall, the three-body structure is closely related to the average number of hydrogen bonds per water molecule reported in \autoref{tab:hbonds}. 
The over-structured ML-FFs generally predict a significantly larger number of hydrogen bonds per molecule, consistent with their more strongly developed local water structure. 
The main exception is RPBE, which exhibits pronounced tetrahedral ordering despite a comparatively small number of hydrogen bonds per molecule. 
This apparent discrepancy can be attributed to its lower liquid density, which reduces the number of neighboring molecules available for hydrogen bond formation while still favoring a strongly tetrahedral arrangement of the local environment.
\begin{table}
\centering
\caption{Average number of hydrogen bonds per water molecule obtained from a geometric criterion with $r_{\mathrm{OO}} < 3.5$ Å and $\angle\mathrm{O-H\cdots O} > 150^\circ$.}
\label{tab:hbonds}
\begin{tabular}{lcc}
\toprule
Model & H-bonds per water \\
\midrule
SPC/E & 2.94 $\pm$ 0.06 \\
PBE & 3.72 $\pm$ 0.03 \\
PBE-D3 & 3.54 $\pm$ 0.04 \\
PBE-TS & 3.62 $\pm$ 0.04 \\
R2SCAN+rVV10 & 3.16 $\pm$ 0.05 \\
RPBE & 2.87 $\pm$ 0.07 \\
RPBE-D3 & 2.98 $\pm$ 0.05 \\
vdW-DF-cx & 3.45 $\pm$ 0.05 \\
\bottomrule
\end{tabular}
\end{table}

\subsection*{Coupled Translational and Orientational Excess Entropy}
\begin{figure}
    \centering
    \includegraphics[width=1\linewidth]{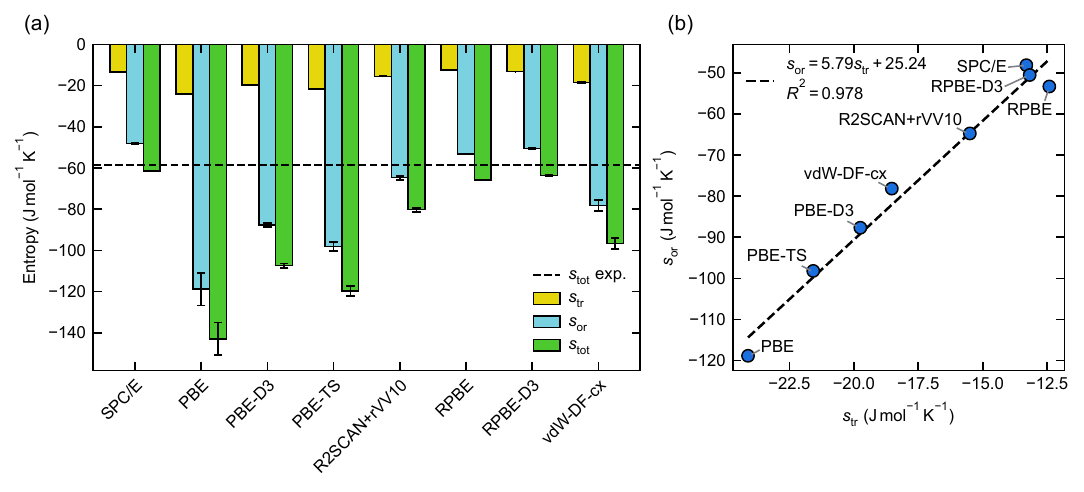}
    \caption{(a) Comparison of the translational $s_\mathrm{tr}$, orientational $s_\mathrm{or}$ and total $s_\mathrm{tot}$ entropy from PCFs of the classical SPC/E and the ML-FFs. Here, the dashed horizontal line gives the experimental value\autocite{wagner_iapws_2000} at $T=\SI{300}{\kelvin}$ and $\rho = \SI{0.997}{\kilogram\per\liter}$. (b) Relationship between $s_\mathrm{or}$ and $s_\mathrm{tr}$ across the investigated force fields. The dashed line shows a linear fit to the data.}
    \label{fig:Entropy}
\end{figure}

Having established how the different force fields modify the molecular structure of liquid water, we next quantify these structural changes in terms of the excess entropy and examine how its translational and orientational contributions are related. 
Both contributions are computed directly from the corresponding molecular pair correlations.
The computed values of the translational $s_\mathrm{tr}$, orientational $s_\mathrm{or}$, and total excess entropy $s_\mathrm{tot}$ for the classical SPC/E model and the \emph{ab initio}-derived ML-FFs are demonstrated in \autoref{fig:Entropy}(a).
As expected for a polar liquid, the orientational contribution to the total excess entropy is significantly larger than the translational contribution independent of the XC-functionals. 
The total excess entropies of SPC/E and RPBE-D3 are \SI{-61.42(26)}{\joule\per\mol\per\kelvin} and \SI{-63.68(47)}{\joule\per\mol\per\kelvin}, respectively, and therefore in close agreement to the experimental value\autocite{wagner_iapws_2000} of \SI{-58.53}{\joule\per\mol\per\kelvin}. 
For the RPBE ML-FF, the agreement with experiment should be interpreted with caution, as it appears to arise from a cancellation of errors between an underestimated density and over-structured RDF and ODFs.
All the other XC-functionals yield significantly smaller excess entropy values, in line with the observed translational and orientational over-structuring of the fluid. 

Interestingly, the force field  dependent variations in $s_{\mathrm{tr}}$ and $s_{\mathrm{or}}$ are not independent; rather, they follow a consistent trend across all investigated water models. 
As illustrated in \autoref{fig:Entropy}(b), $s_{\mathrm{or}}$ scales linearly with $s_{\mathrm{tr}}$, suggesting that increased structural ordering in the translational distribution necessitates a corresponding increase in orientational correlations. 
This linear relationship implies that translational excess entropy alone is an effective descriptor for the structure of water. 
Notably, its easy accessibility from the RDF makes it a practical tool for structural analysis, even when employing high-level ab initio methods, without the necessity to resolve the full six-dimensional PCF.

\subsection*{Transport Properties Follow the Entropic Ordering}
\begin{figure}
    \centering
    \includegraphics[width=0.5\linewidth]{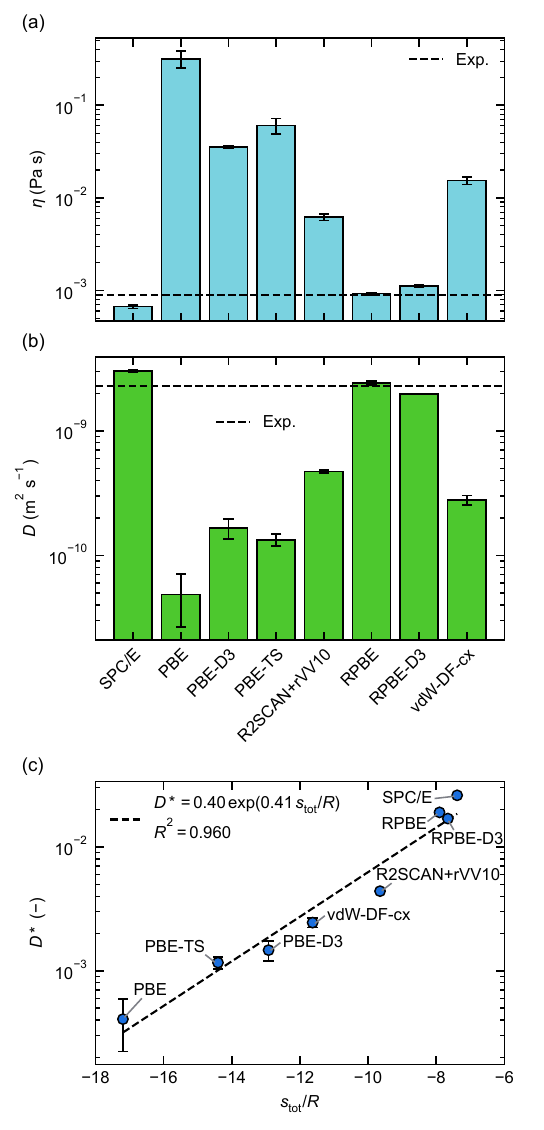}
    \caption{Comparison of (a) the viscosity $\eta$ and (b) the self-diffusion coefficient $D$ for the classical SPC/E and the ML-FFs based on different XC-functionals. The dashed horizontal lines give the respective experimental values\autocite{korsonViscosity1969,holzTemperaturedependent2000} at $T=\SI{298.15}{\kelvin}$. (c) Reduced self-diffusion coefficient $D^*$ as a function of the total excess entropy $s_\mathrm{tot}$ across the investigated force fields. The dashed line shows an exponential fit to the data.}
    \label{fig:Transport}
\end{figure}

Next, we analyze the dynamic properties of the different water ML-FFs and SPC/E and connect them to the inherent molecular structure. The results for the dynamic viscosity $\eta$ and the finite-size corrected self-diffusion coefficient $D$ are shown in \autoref{fig:Transport}(a) and (b), respectively.
First and foremost, the inverse relationship between $D$ and $\eta$, consistent with the Stokes-Einstein relation\autocite{einstein_uber_1905}, is clearly apparent, with ML-FFs exhibiting higher $\eta$ yielding lower $D$, and vice versa.
The SPC/E model under-predicts the viscosity and over-predicts the diffusion coefficient as expected \autocite{tazi_diffusion_2012,celebi_role_2019}, yielding $\eta = \SI{0.667(2.8)}{\milli\pascal\second}$ and $D = \SI{3.02(0.0548)e-9}{\meter\squared\per\second}$, compared to the experimental values of $\eta_\mathrm{exp} = \SI{0.890}{\milli\pascal\second}$\autocite{korsonViscosity1969} and $D_\mathrm{exp} = \SI{2.3e-9}{\meter\squared\per\second}$\autocite{holzTemperaturedependent2000}, respectively.
In contrast, the ML-FF that best reproduces the RDFs and entropy, RPBE-D3, over-predicts the viscosity and under-predicts the self-diffusion coefficient by approximately \SI{25}{\percent}, with $\eta = \SI{1.113(0.026)}{\milli\pascal\second}$ and $D = \SI{1.919(0.095)e-9}{\meter\squared\per\second}$, respectively. 
The seemingly better agreement of RPBE without dispersion correction arises from error cancellation. RPBE underestimates the density by more than \SI{20}{\percent}, resulting in less dense packing, leading to fewer constraints on molecular motion. This results in a lower viscosity and higher self-diffusion coefficient, although the structural analysis in \autoref{fig:RDF-OO}, \autoref{fig:ODF}, and \autoref{fig:ThreeBodyAngle} would otherwise suggest more restricted molecular mobility.
Conversely, the over-structuring observed for PBE, together with a density prediction closer to the experimental value, leads to a substantially higher viscosity, as individual water molecules face a larger barrier to leaving their tetrahedral coordination environment. Consistently, the self-diffusion coefficient is drastically underestimated. While dispersion corrections improve the transport behavior, the experimental values  remain out of reach by more than an order of magnitude. R2SCAN+rVV10 fails to reproduce the second hydration shell, and its transport properties are consequently also unsatisfactory. Finally, for vdW-DF-cx, with a high calculated entropy, the viscosity and diffusion coefficients are also over- and under-predicted, respectively. 

This behavior is consistent with excess-entropy scaling,\autocite{rosenfeld_excess-entropy_2000,dyre_perspective_2018} according to which reduced transport coefficients such as the diffusivity $D^*$ and viscosity $\eta^*$ are approximately governed by the excess entropy of the liquid. The reduced diffusivities were therefore fitted to
\begin{equation}
D^* = A \exp\left(\beta \frac{s_\mathrm{tot}}{R}\right) ,
\end{equation}
where $A$ and $\beta$ are empirical fitting parameters and $s_\mathrm{tot}/R$ is the molar excess entropy scaled by the universal gas constant. 
Here, the reduced diffusivity is defined as
\begin{equation}
D^* = D \rho^{1/3}
\left(
\frac{m}{R T}
\right)^{1/2},
\end{equation}
where $\rho$ is the number density and $m$ is the molar mass of water molecule. 

Within this framework, over-structured water models yield a more negative excess entropy and therefore tend to underestimate diffusion and overestimate viscosity, whereas more disordered hydrogen bond networks are associated with enhanced molecular mobility. As shown in \autoref{fig:Transport}(c), the calculated self-diffusion coefficient follow this excess-entropy trend. This agreement confirms that the differences in transport between the force fields are directly linked to the underlying degree of molecular ordering encoded in the excess entropy.

\subsection*{Classical and ML-FFs Converge through Effective Interactions}
Finally, it is important to understand why the classical SPC/E model and the RPBE-D3 ML-FF exhibit remarkably similar behavior and are both in good agreement with experiment in terms of density, structure, excess entropy, and, consequently, diffusivity and viscosity.
This raises the question of why two conceptually different models lead to such similar predictions. 
At first sight, this similarity is not obvious. 
Since the structural ordering is influenced by long-range electrostatic interactions, we investigated the partial charges predicted by the different XC-functionals in the water molecules with Bader and Born effective charge analysis.
The Bader charges predicted for RPBE-D3 are more than \SI{40}{\percent} larger than the fixed partial charges used in the SPC/E force field, see \autoref{tab:partial_charge}. 
However, Bader charges primarily reflect a static partitioning of the electronic density and should therefore not be interpreted directly as effective interaction charges. 
A more relevant measure of the electrostatic response is provided by the Born effective charges, which quantify the change in polarization upon atomic displacement. Interestingly, the isotropic Born effective charges obtained for RPBE-D3 deviate less than \SI{5}{\percent} from the partial charges used in SPC/E. 
In a simplified picture, this suggests that the effective electrostatic interactions between atoms are rather similar in both models. A second important connection arises from the treatment of dispersion interactions. 
Morawietz \textit{et al.}\autocite{morawietz_how_2016} showed that van-der-Waals interactions are crucial for accurately describing the density anomaly of water and the flexibility of its hydrogen bond network. 
In RPBE-D3, dispersion is included through the D3 correction\autocite{grimme2010},
\begin{equation}
E_{\mathrm{disp}} =
-\sum_{A>B}
\left[
s_6
\frac{C_6^{AB}}{R_{AB}^{6}}
f_{d,6}(R_{AB})
+
s_8
\frac{C_8^{AB}}{R_{AB}^{8}}
f_{d,8}(R_{AB})
\right]
\end{equation}
where $R_{AB}$ is the distance between atoms $A$ and $B$, $C_6^{AB}$ and $C_8^{AB}$ are environment-dependent dispersion coefficients, $s_6$ and $s_8$ are functional-specific scaling parameters, and $f_{d,n}$ denotes the short-range damping function.
Importantly, the leading $R_{AB}^{-6}$ term has the same dependence on the distance between atoms $A$ and $B$ as the attractive dispersion contribution in SPC/E\autocite{berendsen_missing_1987}, reflecting the long-range electronic dispersion that arises from correlated fluctuations of the electron density. 
The D3 correction also contains an $R_{AB}^{-8}$ contribution and damping functions, but previous works show that RPBE-D3 with zero damping gives better results for water than the Becke-Jonson damping.\autocite{sakong_structure_2016}
The dominant long-range attraction is therefore described in a comparable way. 
Thus, despite their different construction, SPC/E and RPBE-D3 share similar effective electrostatic interactions and comparable long-range dispersion physics. 
Consequently, this provides a plausible explanation for why both models reproduce the density, structure, entropy, and transport properties of liquid water with comparable accuracy.

\begin{table}[h]
	\captionof{table}{Bader charge and isotropic Born effective charge analysis expressed in units of the elementary charge. For comparison, the fixed partial charges of H and O in the SPC/E model\autocite{berendsen_missing_1987} are \num{+0.4238} and \num{-0.8476}, respectively.} \label{tab:partial_charge}
	\centering
		\begin{tabular}{lcccc}\toprule[1.2pt]	
        & \multicolumn{2}{c}{Bader charge} & \multicolumn{2}{c}{Born effective charge}  \\\cmidrule(lr){2-3}  \cmidrule(lr){4-5}
			XC-functional &  H     & O    & H & O   \\\midrule
            PBE	           & 0.617	$\pm$0.023	& -1.234 $\pm$0.034 &0.483 $\pm$0.082&-0.967 $\pm$0.101 \\
            PBE-D3	       & 0.619 $\pm$0.020	& -1.239 $\pm$0.029 &0.511 $\pm$0.085&-1.022 $\pm$0.103 \\
            PBE-TS	       & 0.621	$\pm$0.020	& -1.241 $\pm$0.029 &0.526 $\pm$0.087&-1.052 $\pm$0.111 \\
            R2SCAN+rVV10   & 0.644	$\pm$0.024	& -1.287 $\pm$0.036 &0.504 $\pm$0.097& -0.995 $\pm$0.129 \\
            RPBE	       & 0.590	$\pm$0.024	& -1.180 $\pm$0.036 &0.366 $\pm$0.070&-0.731 $\pm$0.077 \\
            RPBE-D3	       & 0.597	$\pm$0.022	& -1.195 $\pm$0.033 &0.430 $\pm$0.080&-0.860 $\pm$0.092 \\
            vdW-DF-cx	   & 0.616	$\pm$0.023	& -1.232 $\pm$0.039 &0.496 $\pm$0.097&-0.992 $\pm$0.122 \\\bottomrule
		\end{tabular}
\end{table}

\section*{Conclusion}
Our simulations demonstrate that the molecular structure of water is critically dependent on the choice of XC-functional employed during ML-FF training. 
While all tested functionals lead to a characteristic hydrogen bond network, the omission of long-range dispersion interactions induces severe over-structuring. This is evidenced by the radial and orientational pair distribution functions, as well as the three-body structural properties. 
Specifically, the heightened translational and orientational ordering promotes an ice-like arrangement, where the predominantly tetrahedral network is disrupted by only a minimal fraction of interstitial water molecules. 
This structural intensification leads to a significant underestimation of the excess entropy.

We find that the RPBE-D3 ML-FF provides the best agreement with the experimental density, radial distribution function, excess entropy, viscosity, and self-diffusivity among 7 different XC-functionals.
Across all studied models, orientational excess entropy remains the dominant component of the total excess entropy. 
Notably, we identify a linear relationship between orientational and translational excess entropy, suggesting that the more readily accessible translational component can serve as a reliable proxy for assessing the quality of ML-FFs. 
Furthermore, these structural discrepancies directly propagate to the predicted self-diffusion coefficients and viscosity, following the excess entropy scaling law observed in other liquids. 
Error cancellation, specifically the compensation of over-structuring by lower density, can yield seemingly accurate entropy and transport coefficients, but such results may be physically misleading. 

Despite their conceptual difference, the classical SPC/E and RPBE-D3 ML-FF give very similar results, which can be explained by their similar descriptions of the effective electrostatic and long-range dispersion physics.
Our results indicate that effective electrostatics of the ML-FF are better captured through Born-effective charges rather than a static Bader-charge analysis, owing to a more accurate description of the polarization change induced by atomic displacement. 
Ultimately, through the joint analysis of molecular structure, excess entropy, transport properties, and effective charges, we establish a unified physical picture of how the underlying XC-functional governs the thermodynamic and dynamical behavior of water. 

\section*{Computational Details}
\subsection*{Machine learning force fields}
Ab initio simulations were performed with the Vienna Ab Initio Simulation Package (VASP) \autocite{kresse1996,kresse1996a} with projector-augmented-wave (PAW) potentials \autocite{kresse1999,blochl1994} and a $\Gamma$-centered Monkhorst-Pack $2\times2\times2$ k-mesh \autocite{pack1977}. We used standard pseudopotentials for H and O with a cut-off energy of \SI{700}{eV}. We compare 7 different combinations of exchange-correlation (XC) functionals and dispersion corrections: PBE \autocite{perdew1996}, PBE-D3 with zero damping \autocite{perdew1996,grimme2010}, PBE-TS \autocite{perdew1996,tkatchenko2009}, RPBE \autocite{hammer1999}, RPBE-D3 with zero damping \autocite{hammer1999,grimme2010}, R2SCAN+rVV10 \autocite{ning2022}, and vdW-DF-cx \autocite{berland2014}. ML-FFs were trained for each XC-functional using VASPs on-the-fly-training scheme during NpT MD simulations at \SI{1}{bar} with 64 water molecules and 50000 time-steps of 1 fs. The threshold for electronic convergence was set to $<\SI{1E-6}{eV}$ in energy change. For training only, we set the mass of H to \SI{8}{amu} to reduce high-frequency oscillations. The temperature was increased from 100 to \SI{400}{K} along the \SI{50}{ps} trajectory using a Langevin thermostat with friction coefficients of \SI{10}{ps^{-1}} for the atoms and \SI{3}{ps^{-1}} for the lattice. The maximum number of basis functions in the kernel was set to 6000, we used element reduced descriptors, and scaled the total energy to the energy of the isolated atoms. Especially the last setting was crucial to prevent bond-breaking in the ML-FF simulations. We used the default hyperparameters provided by VASP for refitting according to the VASP procedure.

\subsection*{Molecular Dynamics Simulations}
Large-scale molecular dynamics simulations with the ML-FFs were carried out using the VASP MD engine in simulation boxes containing 512 water molecules. 
The simulations were performed at \SI{300}{K} with a time step of \SI{0.5}{fs} using physical atomic masses.
Production simulations were carried out in the NVT ensemble at densities obtained from prior \SI{1}{ns} NpT equilibration runs at \SI{1}{bar} and \SI{300}{K}. The temperature in the NVT ensemble was maintained using a Nose-Hoover thermostat with a Nosé-mass of 5 in VASP and \SI{0.1}{\pico\second} damping parameter in LAMMPS. We chose a different thermostat for the NVT simulations, as the Langevin thermostat is good for phase space sampling, but introduces dynamical artifacts that alter the molecular motion.\autocite{basconiEffects2013}  
The SPC/E simulations were performed in LAMMPS\autocite{thompson_lammps_2022} using parameters consistent with those employed in the ML-FF simulations.
Additionally, the intramolecular water geometry was constrained using the SHAKE algorithm.\autocite{ryckaert_numerical_1977}

\subsection*{Pair-correlation functions}
To characterize the molecular structure predicted by the ML-FFs, the MD trajectories were analyzed using the full six-dimensional molecular pair correlation function\autocite{lazaridisOrientational1996} (PCF), $g(r,\bm{\omega})$, defined in \autoref{fig:Workflow_Definition}. Here, $r$ denotes the oxygen--oxygen separation, while $\bm{\omega}$ comprises the five relative orientational degrees of freedom of a molecular pair. The molecular dipole vectors are denoted by $\bm{D}_1$ and $\bm{D}_2$, and the intermolecular axis is defined as $\bm{r}=\overrightarrow{O_1O_2}$, where $O_1$ and $O_2$ are the oxygen atoms of the respective water molecules. The relative orientations are described by $\cos\theta_1=\hat{\bm{D}}_1\cdot\hat{\bm{r}}$, $
\cos\theta_2=\hat{\bm{D}}_2\cdot(-\hat{\bm{r}})$, $\cos\phi=(\hat{\bm{r}}\times\hat{\bm{D}}_1)\cdot[\hat{\bm{D}}_2\times(-\hat{\bm{r}})]$, $\cos\chi_1=\widehat{\bm{H}_{11}\bm{H}_{12}}\cdot(\hat{\bm{r}}\times\hat{\bm{D}}_1)$, and $\cos\chi_2=\widehat{\bm{H}_{21}\bm{H}_{22}}\cdot[\hat{\bm{D}}_2\times(-\hat{\bm{r}})]$. The two hydrogen atoms within each water molecule are labeled arbitrarily.
The molecular PCF can be decomposed into a translational contribution, represented by the oxygen--oxygen radial distribution function $g_\mathrm{OO}(r)$, and a conditional orientational contribution $g(\bm{\omega}\mid r)$: 
\begin{equation}
g(r,\bm{\omega})=g_\mathrm{OO}(r),g(\bm{\omega}\mid r) ~ .
\end{equation}
The first factor resembles the standard oxygen-oxygen (O-O) radial distribution function (RDF), 
\begin{equation}
g_\mathrm{OO} (r) = \frac{1}{\Omega} \int g(r,\bm{\omega}) \dd{\bm{\omega}} \, ,
\end{equation}
where $\Omega$ is the integral over the angles $\bm{\omega}$.
The second factor, $g(\bm{\omega}|r)$, is the orientational distribution function (ODF) for the relative orientation of two molecules at the distance $r$, which is normalized such that
\begin{equation}
\int g(\bm{\omega}|r) \dd{\bm{\omega}} = \Omega \, .
\end{equation}
For the comparison of the orientational structure, the ODF was decomposed into its marginal ODFs obtained from
\begin{equation}
g_i(\omega_i) = \frac{\int g(\bm{\omega}) \,\dd{\bm{\omega}_{j \neq i}}}{\int \dd{\bm{\omega}_{j \neq i}}}\,,
\end{equation}
where $\omega_i$ denotes one of the angular coordinates and $\bm{\omega}_{j\neq i}$ represents the set of all remaining orientational degrees of freedom. The radial dependence was removed by averaging the orientational contributions over all molecular pairs within the first hydration shell, $r \leq R_1$, where $R_1$ denotes the position of the first minimum in the oxygen--oxygen RDF.

For the full six-dimensional PCF, also used in the subsequent entropy calculation, the radial coordinate was discretized with a bin width of $\Delta r = \SI{0.1}{\angstrom}$, while each angular coordinate was discretized using bins of $\Delta \omega =  \SI{10}{\degree}$. 
A separate calculation of the oxygen--oxygen, oxygen--hydrogen and hydrogen--hydrogen RDFs, used only for structural comparison and not for the entropy calculation, was performed with a finer radial bin width of \SI{0.02}{\angstrom}. 
The coarser binning of the full PCF was required to ensure sufficient sampling of the multidimensional distribution. 
The PCFs were obtained from \num{100000} snapshots sampled at intervals of \SI{10}{\femto\second}.

\subsection*{Three-body structure}
Three-body angle distributions and the corresponding tetrahedral order parameter $q$ were evaluated from the same trajectories used for the pair-correlation analysis. 
For each water molecule, the four nearest oxygen neighbors were identified and used to compute the angles $\alpha_{jk}$ between the vectors connecting the central molecule to neighbors $j$ and $k$. 
The orientational order parameter was then calculated as\autocite{errington_relationship_2001}
\begin{equation}
q = 1 - \frac{3}{8} \sum_{j=1}^{3} \sum_{k=j+1}^{4} \left( \cos\alpha_{jk} + \frac{1}{3} \right)^2 ~.
\end{equation}
Here, $q$ ranges from \num{-3} to \num{1}, with $q=0$ corresponding to an uncorrelated arrangement, while $q=1$ corresponds to a perfectly tetrahedral local environment. 
The three-body angle and $q$-factor distributions were discretized into 180 bins, corresponding to bin widths of $\Delta \alpha = \SI{1}{\degree}$ and $\Delta q \approx \num{0.022}$, respectively.

\subsection*{Excess entropy}
To quantify the structural order of the ML-FF water models, the excess entropy was calculated from the molecular PCFs following Lazaridis and Karplus,\autocite{lazaridisOrientational1996} and Giuffrè et al. \autocite{giuffreEntropy2010} The total pair-correlation entropy, $s_\mathrm{tot}$, was resolved into translational and orientational contributions,
\begin{equation}
s_\mathrm{tot} = s_\mathrm{tr} + s_\mathrm{or} ~ .
\end{equation}
The translational excess entropy solely depends on the oxygen-oxygen RDF $g_\mathrm{OO}$ and is given by 
\begin{equation}
s_\mathrm{tr} = - 2 \pi \, k_\mathrm{B} \, \rho \int_0^\infty \left[ g_\mathrm{OO}(r) \, \ln{g_\mathrm{OO}(r) - g_\mathrm{OO}(r) + 1} \right] r^2 \dd{r} ~ ,
\end{equation}
where $k_\mathrm{B}$ denotes Boltzmann constant and $\rho$ the computed bulk density. 
The orientational excess entropy $s_\mathrm{or}$ follows from
\begin{equation}
\label{eq:s_or_def}
s_\mathrm{or} = - 2 \pi \,  k_\mathrm{B}  \rho \int_0^\infty g_\mathrm{OO}(r) \, \mathcal{S}_\mathrm{or}(r) \, r^2 \dd{r} \, ,
\end{equation}
defined in terms of the local orientational excess entropy 
\begin{equation}
\mathcal{S}_\mathrm{or}(r) = \frac{1}{\Omega} \int g(\bm{\omega}|r) \, \ln{g(\bm{\omega}|r)} \dd{\bm{\omega}} ~ .
\end{equation}
For the production runs used to evaluate the PCFs for the entropy calculation, snapshots were collected every \SI{10}{\femto\second} from three independent NVT trajectories, with \num{100000} snapshots sampled from each trajectory.
To reduce finite-sampling effects in the entropy calculation\autocite{giuffreEntropy2010}, the entropy was extrapolated to the infinite-sampling limit, as described in detail in the Supplementary Information. 
The reported excess entropy values correspond to the averages over the three independent trajectories.

\subsection*{Transport properties}
The viscosity and self-diffusion coefficient were calculated from five independent NVT production trajectories for each ML-FF, with a total production time of \SI{1}{ns} each. Atomic positions used to determine the self-diffusion coefficient and the components of the stress tensor used to calculate the viscosity were sampled every 2 and 100 time steps, respectively. 

The viscosity was computed using a modified Green-Kubo relation in which the autocorrelation functions of all independent components of the symmetric traceless stress tensor are averaged to improve statistical convergence,\autocite{maginn_best-practices_2018, daivis_viscosity_1994, mondello_viscosity_1997}
\begin{equation}
\eta(t)
=
\frac{V}{10 k_\mathrm{B} T}
\sum_{\alpha,\beta}
\int_0^t
\left\langle
P_{\alpha\beta}(t_0+\tau)
P_{\alpha\beta}(t_0)
\right\rangle_{t_0}
\dd{\tau} ~ .
\end{equation}
Here, $\eta(t)$ is the cumulative viscosity evaluated up to the correlation time $t$, $V$ is the volume of the simulation cell, and $\langle\cdots\rangle_{t_0}$ denotes an average over multiple time origins $t_0$.
The tensor $P_{\alpha\beta}$ corresponds to the symmetric traceless part of the stress tensor $\sigma_{\alpha\beta}$ and is defined as
\begin{equation}
P_{\alpha\beta}
=
\frac{\sigma_{\alpha\beta}+\sigma_{\beta\alpha}}{2}
-
\frac{\delta_{\alpha\beta}}{3}
\sum_\gamma \sigma_{\gamma\gamma} ~ ,
\end{equation}
where $\delta_{\alpha\beta}$ is the Kronecker delta and $\alpha,\beta,\gamma \in \qty{x,y,z}$.
The stress autocorrelation functions were evaluated using all available time origins and truncated at a maximum lag time selected to retain the statistically converged portion of the correlation function. 
The resulting cumulative viscosity was fitted using a biexponential function,\autocite{rey-castro_transport-properties_2006}
\begin{equation}
\eta(t)
=
A\alpha\tau_1
\left(1-e^{-t/\tau_1}\right)
+
A(1-\alpha)\tau_2
\left(1-e^{-t/\tau_2}\right) ~ ,
\end{equation}
where $A$, $\alpha$, $\tau_1$, and $\tau_2$ are fitting parameters. 
Extrapolation of the fitted function to infinite correlation time yields the final viscosity reported in the results,
\begin{equation}
\eta
=
\lim_{t\to\infty}\eta(t)
=
A\alpha\tau_1
+
A(1-\alpha)\tau_2 ~ .
\end{equation}

The self-diffusivity was calculated using the Einstein relation \autocite{allen_computer-simulation_2017}
\begin{equation}
    D_\mathrm{PBC} = \frac{1}{6N} \lim_{t\to\infty} \dv{t} \sum_{i=1}^N \left\langle \lvert\bm{r}_i(t_0 + t) - \bm{r}_i(t_0)\rvert^2 \right\rangle_{t_0} ~ ,
\end{equation}
where $D_\mathrm{PBC}$ denotes the self-diffusion coefficient obtained for the finite simulation cell under periodic boundary conditions, $N$ is the number of molecules and $\bm{r}_i$ the unwrapped center-of-mass position of the $i$-th molecule.
The mean-squared displacement (MSD) was calculated by averaging over all available time origins. 
A linear regression was then performed over the time interval corresponding to the diffusive regime, identified by a slope of approximately 1 in a log–log plot of the MSD versus time.
The computed self-diffusion coefficients were corrected for finite-size effects using the Yeh–Hummer correction.\autocite{yeh_finite-size-correction_2004} These effects arise from the long-range hydrodynamic interactions between a diffusing molecule and its periodic images, which artificially reduce molecular mobility in finite simulation cells.\autocite{celebi_finite-size_2021} The corrected self-diffusion coefficient is therefore given by
\begin{equation}
    D = D_\mathrm{PBC} + \frac{k_\mathrm{B} T \xi}{6\pi\eta L} ~,
\end{equation}
where $D$ is the final self-diffusion coefficient reported in the results, $\xi = 2.837297$ a dimensionless constant and $L$ the length of the cubic simulation cell.

Detailed plots illustrating the fitting procedures and analyses used to determine the self-diffusion coefficient and viscosity are provided in the Supporting Information.

\section*{Acknowledgments}
This research was funded in whole by the Austrian Science Fund (FWF) [10.55776/COE5] (Cluster of Excellence MECS). For open access purposes, the author has applied a CC-BY public copyright license to any author accepted manuscript version arising from this submission. Computational results were achieved using the Austrian Scientific Computing (ASC) infrastructure.

We thank Georg Kresse and his group at the University of Vienna, in particular Angela Rittsteuer and Michael Sahre, for helpful discussions on Born effective charges.

\section*{Supporting Information}

The Supporting Information contains further details on the electronic structure, the machine learning procedure, and the structural analysis.

\section*{Author contributions}
A.K. developed the machine-learning models and performed the ML-FF simulations. F.A. analyzed the structural and entropic properties, and led the writing of the manuscript. N.N. analyzed the transport properties. A.T.C. contributed to the conceptualization of the work and provided guidance on the theoretical analysis and atomistic simulations. M.V. acquired funding and supervised the project. All authors discussed the results, reviewed the manuscript, and contributed to its final revision.

\section*{Data and Software Availability}
The data underlying this study are available in the published article, the Supporting Information, and through the TU Wien Research Data Repository \autocite{altmann_local_2026}. The deposited dataset also includes the scripts and code used to process the data and reproduce the analyses.

\printbibliography

\end{mainmatter}
\end{document}